\documentstyle[amssymb,amsmath,aps,pra]{revtex}

\newtheorem{defi}{Definition}
\newtheorem{lemma}[defi]{Lemma}
\newtheorem{thm}[defi]{Theorem}

\newtheorem{prop}[defi]{Proposition}

\newcommand{\qed}{\hfill $\Box$}
\newcommand{\tr}{{\operatorname{Tr}\,}}

\newcommand{\bra}[1]{{\langle{#1}|}}
\newcommand{\ket}[1]{{|{#1}\rangle}}

\newcommand{\E}{{\mathbb{E}}}
\newcommand{\1}{{\openone}}

\newlength{\blank}
\newlength{\equalsign}
\newenvironment{beweis}[1]
{{\noindent\emph{Proof\hspace{\blank}{#1}.\ }}}
{\hfill $\Box$\vskip 0.5\baselineskip}

\begin{document}

\twocolumn
\narrowtext

\title{Compression of quantum measurement operations}
\author{Andreas 
Winter\thanks{Email: \texttt{winter@mathematik.uni-bielefeld.de}}}
\address{SFB 343, Fakult\"at f\"ur Mathematik, Universit\"at Bielefeld,\\
Postfach 100131, D--33501 Bielefeld, Germany}
\author{Serge Massar\thanks{Email: \texttt{smassar@ulb.ac.be}}}
\address{Service de Physique Th\'{e}orique, Universit\'{e} 
Libre de Bruxelles,\\
CP 225, Boulevard de Triomphe, B--1050 Bruxelles, Belgium}
\date{December 22, 2000}

\maketitle


\begin{abstract}
We generalize recent work of Massar and Popescu dealing with the amount of 
classical data that is produced by a quantum measurement on a quantum
state ensemble. In the previous work it was shown that quantum measurements
generally contain spurious randomness in the outcomes and that this
spurious randomness 
can be eliminated by carrying out collective measurements on many independent
copies of the system. In particular it was shown that, without
decreasing the amount of knowledge the measurement provides about the
quantum state,
one can always reduce the 
amount of data produced by the measurement to the entropy 
$H(\rho)=-\tr\rho\log\rho$ of the ensemble.
\par
Here we extend this result by giving a more
refined description of what constitute equivalent measurements (that
is measurements which provide the same knowledge about the quantum
state)
and also by considering incomplete measurements. In particular we show that
one can always associate to a POVM with elements $a_j$, an equivalent
POVM acting on many independent copies of the system which produces an 
amount of data asymptotically equal to the entropy defect of an ensemble
canonically associated to the ensemble average state $\rho$ and the 
initial measurement $(a_j)$.
In the case where the measurement is not maximally refined this amount 
of data is strictly less than the amount $H(\rho)$ obtained in
the previous work. This result is obtained by
a novel technique to analyze random selections.
We also show that this is the best achievable,
i.e. it is impossible to devise a
measurement equivalent to the initial measurement 
$(a_j)$ that produces less data.
\par
We discuss the interpretation of these
results. In particular we show how they  
can be used to provide a precise and model independent 
measure of the amount of knowledge that is obtained about a quantum
state by a quantum measurement. We also discuss in detail the relation
between our results and Holevo's bound, at the
same time providing a new proof of this fundamental inequality.
\end{abstract}

\section{Introduction}
  \label{sec:intro}
An essential aspect of quantum mechanics is the measurement
process. Only by measuring can a macroscopic observer obtain
knowledge about a quantum system. 
However the  knowledge that is obtained about the state of a quantum
system is
in general not complete since from the outcome of a measurement
it is in general not possible to infer the initial state. 
Furthermore there are many different
measurements that could be carried out on the system and these
measurements are in general mutually incompatible. 
\par
It is therefore natural  
to try to make measurements as efficient as
possible. The simplest way one can make a measurement efficient is to
devise it in such a way that it provides as much knowledge as possible
about the
state of the system\footnote{In this article we shall distinguish
  between the words ``knowledge'' and ``information''. Thus we shall
  say that a measurement provides knowledge about the state of a
  quantum system, rather than information.
  We introduce this distinction 
  because the second term is often associated with
  the ``mutual information'' between the initial state and the result
  of the measurement. And, as examples show, 
  an efficient measurement is not necessarily 
  one that maximizes the mutual information 
  between the initial state and the result
  of the measurement.}.
This first approach has been extensively studied, see
for instance~\cite{helstrom:detection,holevo:statistical}. 
We note that in some cases it can be
interesting to make an incomplete measurement (which does not provide
maximum knowledge about the system). The incomplete measurement can
then be refined at a later stage by carrying out a second measurement
on the system.
\par
The second way one can make a measurement efficient is to reduce the
amount of classical data it produces. 
Indeed if a measurement produces outcome
$j$ with probability $p_j$, the amount of classical data produced by
the measurement is $I= -\sum_j p_j \log p_j$ (in this
paper $\log$ and $\exp$ are always to base $2$).
This second aspect of
optimizing measurements was first considered
in~\cite{massar:popescu}. 
\par
Minimizing $I$ is interesting for two reasons. First it makes the
measurement less wasteful of resources since it minimizes that
amount of classical data that is produced. Indeed the increase in
entropy --- in the thermodynamic sense --- due to the irreversibility of the 
measurement process will be minimized if the amount of data $I$
produced by the measurement is minimized. Secondly, as argued
in~\cite{massar:popescu},  the minimum value
of $I$ provides a model independent answer to the question
\emph{how much knowledge about a quantum
  system is obtained by a measurement?}
\par
The main result of~\cite{massar:popescu} was to show that it always
possible to reduce $I$ so that it is less or equal than the von
Neumann entropy of the ensemble of quantum states on which the
measurement is carried out. Thus the answer to the above question is  
that a quantum
measurement can provide at most one bit of classical knowledge about 
an unknown qubit. 
\par
However minimizing $I$ is not an easy task. It must be carried
out at the level of the measurement itself and cannot be realized
as a post--processing of the data produced by the measurement. This is
because there are positive operator valued measures (POVM) that provide
maximum knowledge about the state and that have a number of outcomes
that is larger than the von Neumann entropy of the ensemble.
Such measurements add spurious randomness to their outcomes.
To address
this difficulty and remove the spurious randomness
one must define a notion of 
``equivalent'' measurements that yield
the same knowledge about the quantum system and then search among this
class of equivalent measurements for those which minimize the number of
bits $I$ of classical data produced by the measurement. 
It is important to
include in the equivalence classes not only 
measurements on single states, but also 
measurements that act collectively on blocks of independent
states. It is also essential to include in the equivalence class
measurements that differ infinitesimally.
Such extensions are natural in the context of information theory.
We shall refer to the above procedure as 
the ``compression of quantum measurement operations''.
\par
The results of~\cite{massar:popescu} are incomplete in several 
ways and we complete them in the present paper. In particular we 
give a more precise description of
what constitute ``equivalent'' measurements. We then  
obtain lower bounds on the amount of classical data $I$ that
can be produced by equivalent measurements. Finally we construct
measurements that attain the lower bound. 
Both results  apply to general POVMs and in
particular to incomplete measurements (for which the
POVM elements are not all proportional to one dimensional projectors).

\section{Previous results}
\label{sec:previous}
In this section we shall recall the results obtained
in~\cite{massar:popescu}. This will serve as a basis for the
presentation of our new results in the next section.
\par
Consider a quantum ensemble consisting of states $\ket{\psi_i}$ (in
the Hilbert space ${\cal H}$ which we assume to be of finite
dimension $d$ throughout the paper), with probabilities $p_i$
($i=1,\ldots,n$), 
and a measurement POVM ${\bf a}=(a_j)_{j=1,\ldots,m}$. We suppose
that the measurement maximizes a \emph{fidelity}
$$F({\bf a})=\sum_i p_i\sum_j\bra{\psi_i}a_j\ket{\psi_i}F_{ij},$$
where $F_{ij}$ is the contribution (or \emph{gain}) in the case
that on being given state $\ket{\psi_i}$ the POVM hits upon guess
$j$ (which happens with probability $\bra{\psi_i}a_j\ket{\psi_i}$).
Note that this is equivalent to the objective of quantum estimation
theory~\cite{helstrom:detection,holevo:statistical}
to minimize the \emph{cost}. The same minimization problem occurs
in the computation of the so--called \emph{quantum rate distortion function},
as defined in~\cite{bendjaballah:et:al}.
\par
Here the reason for introducing a fidelity is that it allows us to
define in an implicit way a class of equivalent
measurements. Indeed the $F_{ij}$ encode implicitly a property 
about which knowledge can be obtained by a measurement. 
And a measurement that maximizes
$F$ is an optimal measurement for this property. 
One then defines as equivalent all the
measurements that maximize $F$.
\par
It is demonstrated by examples in~\cite{massar:popescu} 
that the number of outcomes $I$  of the optimal measurement can exceed 
the von Neumann
entropy of the ensemble. But it is proved that if a large number 
of independent states are available, then 
one can find an almost optimal measurement that acts collectively on 
all the copies with
logarithm of number of outcomes asymptotically bounded by the von Neumann
entropy of the ensemble. 
This result can be formulated more precisely as follows:
\par  
Introduce the density operators $\rho_i=\ket{\psi_i}\bra{\psi_i}$,
and the average state $\rho=\sum_i p_i\rho_i$. We assume in the sequel
that $\rho>0$ on ${\cal H}$ (otherwise pass to the support of
$\rho$)
and that  ${\cal H}$ is finite dimensional.
Suppose that  a number $l$ of independent states are available. 
The $l$ states are given  by the density operator
$\rho_{i^l}=\rho_{i_1}\otimes\cdots\otimes\rho_{i_l}$, with
probability $p_{i^l}=p_{i_1}\cdots p_{i_l}$. Here and in what
follows $i^l$ is an abbreviation for a tuple $(i_1,\ldots,i_l)$.
\par
The fidelity for the $l$ independent states 
is defined as the sum of the individual fidelities
\begin{equation}
  F_{i^lj^l}=\frac{1}{l}\sum_{k=1}^l F_{i_kj_k}\ .
  \label{multfid}
\end{equation}
This is a crucial aspect of the model: the
fidelity on blocks is constructed from a sum of fidelities on the individual
systems, in fact as the \emph{average} of these fidelities. 
\par
We now consider a POVM ${\bf A}$  on
${\cal H}^{\otimes l}$ and we compute the fidelity for this
POVM. This POVM has $M$ outcomes labeled by
$\mu=1,\ldots,M$. In order to compute the 
block fidelity  (\ref{multfid})
we must associate to each POVM outcome $\mu$ a tuple of guesses
$j^l_\mu$. Hence the individual POVM elements will be denoted
$A_{j^l_\mu}$. 
\par
One possible example is the product POVM 
${\bf a}^{\otimes l}$ which consists of all operators
$a_{j^l}=a_{j_1}\otimes\cdots\otimes a_{j_l}$.  One easily checks
that in this case the fidelity on blocks
$$F({\bf a}^{\otimes l})=
     \sum_{i^l} p_{i^l}\sum_{j^l} 
     \bra{\psi_{i^l}}a_{j^l}\ket{\psi_{i^l}}F_{i^lj^l}$$
equals the single letter fidelity $F({\bf a})$. In this case the
number of outcomes is equal to the maximum possible number of
guesses, $m^l$.
However in general the number of guesses $M$ 
may be smaller than the number of possible tuples. Thus there can be 
some tuples that are never associated with a POVM element, and hence 
can never constitute a guess. However even when $M$ is less than the
number of possible guesses, we can still compute the average
fidelity.
\par
What we are after is a POVM ${\bf A}=(A_{j^l_\mu})_{\mu=1,\ldots,M}$ on
${\cal H}^{\otimes l}$ whose fidelity $F({\bf A})$
is close to the optimal fidelity
$F_{\text{opt}}$ and with a minimal number $M$
of outcomes. 
This will constitute a POVM
belonging to  the equivalence class for which all the spurious
redundancies have been eliminated.
The central result of~\cite{massar:popescu} is the construction of such a POVM:
\begin{thm}[Massar,~Popescu~\cite{massar:popescu}]
  \label{thm:massar:popescu}
  For $\epsilon>0$ and $l$ large enough there exists a POVM ${\bf A}$
  with fidelity $F({\bf A})\geq F_{\text{opt}}-\epsilon$ and
  $$M\leq\exp(l(H(\rho)+\epsilon))$$
  many outcomes, where $H(\rho)=-\tr\rho\log\rho$ is the von Neumann
  entropy.
  \qed
\end{thm}
\par
We can rewrite the fidelity of ${\bf A}$ as
\begin{equation}\begin{split}
  \label{eq:POVM:fidelity}
  F({\bf A}) &=\sum_{i^l} p_{i^l}\sum_\mu \tr(\rho_{i^l}A_{j^l_\mu})
                                   \frac{1}{l}\sum_{k=1}^l F_{i_kj_{\mu k}} \\
             &=\frac{1}{l}\sum_{k=1}^l\sum_i\sum_j p_i 
                                      \tr(\rho_i A^{(k)}_j) F_{ij},
\end{split}\end{equation}
where (with $[l]=\{1,\ldots,l\}$)
\begin{equation}\begin{split}
  \label{eq:POVM:marginals}
  A^{(k)}_j &=\tr_{\neq k}\left(\left(\rho^{\otimes [l]\setminus k}
                                                  \otimes\1_k\right)
                           \sum_{\mu\text{: }j_{\mu
                               k}=j}A_{j^l_\mu}     \right)   \\
              &=\rho^{-1}\tr_{\neq k}\left(\rho^{\otimes l}
                           \sum_{\mu\text{: }j_{\mu
                               k}=j}A_{j^l_\mu}     \right)   \\
              &\hspace{-.5cm}=\sqrt{\rho^{-1}}\tr_{\neq k}\!
                             \left(\sqrt{\rho}^{\otimes l}
                     \left(\sum_{\mu\text{: }j_{\mu k}=j}A_{j^l_\mu}\right)
                                      \sqrt{\rho}^{\otimes
                                        l}\right)\!\sqrt{\rho^{-1}}. 
\end{split}\end{equation}
To prove the second and third equality recall the defining property of the
partial trace on the composite system ${\cal H}_1\otimes{\cal H}_2$:
$$\forall A\ \ \tr\left(A\tr_2 C\right)=\tr\left((A\otimes\1)C\right).$$
\par
Note that for all $k$ the $A^{(k)}_j$ ($j=1,\ldots,m$) form a POVM
which we shall refer to as \emph{marginals}
of ${\bf A}$. 
The marginals  of ${\bf A}$ describe the action of the  POVM  ${\bf A}$ 
restricted  to the $k$'th state in the block. They
will play a central role in what follows.

\section{Model and main results}
  \label{sec:model}
  Theorem~\ref{thm:massar:popescu} is incomplete in several ways:
  Why do the ensemble states not enter, only their average? Is it
  important that they are pure? Also, what is the deeper reason that
  the fidelity matrix does not enter, nor the structure of the optimal
  measurement? Is the bound on $M$ optimal, or better: under which
  conditions is it optimal? The results below will help clarify these
  questions.
  \par
  We start by analyzing the fidelity constraint on the interesting
  POVMs: this will lead to a series of conditions (C0--C3) of
  increasing strength. Theorem~\ref{thm:massar:popescu} lets us start out from
  the condition
  \begin{equation}
    |F({\bf A})-F({\bf a})|\leq\epsilon.\tag{C0}
  \end{equation}
  Looking again at (\ref{eq:POVM:fidelity}) we observe that $F({\bf A})$ is
  an average over the $l$ positions of equally structured quantities:
  each is an average of the $F_{ij}$, with probabilities
  $p_i\tr(\rho_i A^{(k)}_j)$. Thus, assuming that
  the $|F_{ij}|$ are (without loss of generality) bounded by $1$,
  a POVM ${\bf A}$ on ${\cal H}^{\otimes l}$
  will obtain a fidelity within $\epsilon$ of $F({\bf a})$ (for any
  measurement ${\bf a}$, not only
  the optimal POVM ${\bf a}$ on ${\cal H}$) if, for all $k$,
  the distribution
  $$\left(\frac{1}{l}\sum_{k=1}^l p_i\tr(\rho_i A^{(k)}_j)\right)_{ij}$$
  is close to $(p_i\tr(\rho_i a_j))_{ij}$, i.e.
  \begin{equation}
   \forall k \  \sum_{ij} 
        \left|\left(\frac{1}{l}\sum_{k=1}^l p_i\tr(\rho_i A^{(k)}_j)\right)
                  -p_i\tr(\rho_i a_j)\right|\leq\epsilon.\tag{C1}
  \end{equation}
  This will be satisfied if for each position $k$ and each $i$
  the corresponding sub-terms are close:
  \begin{equation}
    \forall k\forall i\ \sum_j |\tr(\rho_i A^{(k)}_j)-\tr(\rho_i a_j)|
                                      \leq\epsilon.\tag{C2}
  \end{equation}
  And this in turn is satisfied if
  \begin{equation}
    \forall k\ \sum_j \|A^{(k)}_j-a_j\|\leq\epsilon.\tag{C3}
  \end{equation}
  Here the operator sup norm is used. Proof is by the H\"older inequality
  for the trace pairing of operators:
  $$|\tr(AB)|\leq \|A\|_1\cdot\|B\|.$$
  \par
  Now given any ensemble with average state $\rho$ and a POVM
  ${\bf a}=(a_j)_{j=1,\ldots,m}$ a canonical ensemble for $\rho$
  can be written down: the states
  $$\hat{\rho}_j=\frac{1}{\tr(\rho a_j)}{\sqrt{\rho}a_j\sqrt{\rho}},$$
  with probabilities $\lambda_j=\tr(\rho a_j)$.
  \par
  Note that this ensemble has the property that its
  ``square root'' (Holevo~\cite{holevo:sqrt}) or
  ``pretty good'' (Hausladen, Wootters~\cite{hausladen:pgm})
  measurement is exactly ${\bf a}$:
  $$a_j=\sqrt{\rho^{-1}}\lambda_j\hat{\rho}_j\sqrt{\rho^{-1}}.$$
  \begin{thm}
    \label{thm:POVM:compr:1}
    With the above notation and $\epsilon>0$, there exists a POVM
    ${\bf A}=(A_{j^l_\mu})_{\mu=1,\ldots,M}$ with
    $$M\leq
     \exp\left(l\left(H(\rho)-\sum_j \lambda_j 
H(\hat{\rho}_j)\right)+C\sqrt{l}\right)$$
    (where $C$ is a constant depending only on $\epsilon$, 
$d$ and $m$), and such that
    $$\forall k\ \sum_j \|A^{(k)}_j-a_j\|\leq\epsilon.$$
  \end{thm}
  The characteristic constant in the exponent,
  $$I(\lambda;\hat{\rho})=H(\rho)-\sum_j \lambda_j H(\hat{\rho}_j),$$
  is called \emph{entropy defect} of the ensemble
  (Lebedev and Levitin~\cite{levitin:defect}),
  or the \emph{quantum mutual information} 
  between a sender producing letter $j$ with probability $\lambda_j$
  and a receiver getting the letter state $\hat{\rho}_j$
  (see~\cite{holevo:coding}). It is the difference between the
  entropy $H(\rho)$ of the ensemble and its \emph{conditional entropy}
  $H(\hat{\rho}|\lambda)=\sum_j \lambda_j H(\hat{\rho}_j)$.
  \par
  The theorem is in an asymptotic sense best possible:
  \begin{thm}
    \label{thm:POVM:compr:lower}
    Let $0<\epsilon\leq (\lambda_0/2)^2$, with
    $\lambda_0=\min_j \lambda_j$. Then for any POVM
    ${\bf A}=(A_{j^l_\mu})_{\mu=1,\ldots,M}$ such that
    $$\forall k\ \sum_j \|A^{(k)}_j-a_j\|\leq\epsilon,$$
    one has
    $$M\geq \exp\left(l\left(H(\rho)-\sum_j \lambda_j H(\hat{\rho}_j)
       +\frac{3\epsilon}{\lambda_0^2}\log\frac{2\epsilon}
{\lambda_0^2 d}\right)\right).$$
  \end{thm}
\par
These theorems are proven in the following two sections.
They provide
answers to the questions at the beginning of this section. By
demanding a bit more, namely condition C3 instead of the weaker
C0 we find the \emph{optimal} rate of compression for \emph{any}
POVM. This  improves
the previous result (theorem~\ref{thm:massar:popescu})
in all cases where the $a_j$ are not all of rank $1$. This 
optimal compression is independent
of fidelities, as well as independent
of the ensemble structure, except for the average state $\rho$.
\par
These theorems also answer a 
question from~\cite{massar:popescu}, whether the
result of that paper still holds for fidelity measures which cannot
be reduced 
to the form described in the introduction (i.e. an average over certain
fixed numbers, with probabilities $p_i\bra{\psi_i}a_j\ket{\psi_i}$),
e.g. ones which depend in some nonlinear way on the POVM used.
Theorem~\ref{thm:POVM:compr:1} gives an affirmative answer for all
fidelity measures which depend \emph{continuously} on the POVM (to
be precise, on its marginals: the definition of the fidelity on blocks
as average of the single block fidelities seems to remain essential):
an example for this will be discussed in
section~\ref{subsec:holevo} below.

\section{Lower bound}
  \label{sec:lower}
  The proof of theorem~\ref{thm:POVM:compr:lower} rests on some standard
  facts about von Neumann entropy:
  \begin{lemma}
    \label{lemma:data:processing}
    Let $\sigma_j$ be quantum states on ${\cal H}$,
    $\lambda_j$ probabilities,
    and $\sigma=\sum_j \lambda_j\sigma_j$. Then
    $$H(\sigma)\leq H(\lambda)+\sum_j \lambda_j H(\sigma_j).$$
  \end{lemma}
  \begin{proof}
    See~\cite{ahlswede:loeber}:
    this is just the monotonicity of the mutual information
    (data processing inequality)
    under the completely positive and trace preserving map
    $j\mapsto \sigma_j$ from the commutative algebra generated by the $j$
    as mutually orthogonal idempotents to the algebra of linear
    operators on ${\cal H}$.
    \qed
  \end{proof}
  \begin{lemma}
    \label{lemma:I:superadd}
    Let $\sigma_1,\ldots,\sigma_r$ be states on ${\cal H}_1\otimes{\cal H}_2$,
    with probabilities $s_1,\ldots,s_r$, such that
    $\sum_i s_i\sigma_i$ is a product state. Then
    $$I(s;\sigma)\geq I(s;\tr_2\sigma)+I(s;\tr_1\sigma).$$
  \end{lemma}
  \begin{proof}
    This is essentially the sub-additivity of entropy
    (see~\cite{ohya:petz}, p.~23).
    \qed
  \end{proof}
  \begin{lemma}
    \label{lemma:I:grouping}
    Let $\sigma_1,\ldots,\sigma_r$ be states on ${\cal H}$, with
    probabilities $s_1,\ldots,s_r$, and $(J_1,\ldots,J_t)$ a partition 
    of $\{1,\ldots,r\}$. Then, denoting
    $$\tilde{s}_j=\sum_{i\in J_j} s_i\text{ and }
      \tilde{\sigma}_j=\frac{1}{\tilde{s}_j}{\sum_{i\in J_j} s_i\sigma_i},$$
    it follows that
    $$I(s;\sigma)\geq I(\tilde{s};\tilde{\sigma}).$$
  \end{lemma}
  \begin{proof}
    See~\cite{ahlswede:loeber}:
    it is another special case of monotonicity,
    known as \emph{coarse graining}. For a direct proof observe that
    $$\sum_j \tilde{s}_j\tilde{\sigma}_j=\sum_i s_i\sigma_i,$$
    and by the concavity of von Neumann entropy
    $$H(\tilde{\sigma}_j)=H\left(\sum_{i\in I_j} 
\frac{s_i}{\tilde{s}_j}\sigma_i\right)
                         \geq \sum_{i\in I_j} 
\frac{s_i}{\tilde{s}_j}H(\sigma_i).$$
    \qed
  \end{proof}
  \begin{lemma}
    \label{lemma:H:continuity}
    Let $\rho$, $\sigma$ be states on ${\cal H}$, $d=\dim{\cal H}$,
    and $\|\rho-\sigma\|_1\leq\alpha\leq 1/2$. Then
    $$|H(\rho)-H(\sigma)|\leq -\alpha\log\frac{\alpha}{d}.$$
  \end{lemma}
  \begin{proof}
    See~\cite{ohya:petz}, p.~22.
    \qed
  \end{proof}
  \par
  Now we are ready for
  \\
  \begin{beweis}{of theorem~\ref{thm:POVM:compr:lower}}
    On ${\cal H}^{\otimes l}$ consider any POVM
    ${\bf A}=(A_{j^l_\mu})_{\mu=1,\ldots,M}$
    which satisfies the hypothesis of the theorem.
    Then, denoting $\Lambda_\mu=\tr(\rho^{\otimes l}A_{j^l_\mu})$ and
    $$\hat{\rho}^l_\mu=\frac{1}{\Lambda_\mu}
                        \sqrt{\rho}^{\otimes l}
            A_{j^l_\mu}\sqrt{\rho}^{\otimes l},$$
    we find
    \begin{equation*}\begin{split}
      \log M &\geq H(\Lambda)
                \geq H(\rho^{\otimes l})-\sum_\mu 
                         \Lambda_\mu H(\hat{\rho}^l_\mu) \\
             &=    I(\Lambda;\hat{\rho}^l)
                \geq \sum_{k=1}^l I(\Lambda^{(k)};\hat{\rho}^{(k)}),
    \end{split}\end{equation*}
    using lemmas~\ref{lemma:data:processing}, \ref{lemma:I:superadd},
    and \ref{lemma:I:grouping},
    with the marginal distributions given by
    $$\Lambda^{(k)}_j=\tr(\rho A^{(k)}_j)$$
    and the marginal channel states
    $$\hat{\rho}^{(k)}_j=\frac{1}{\Lambda^{(k)}_j}
                             {\sqrt{\rho}A^{(k)}_j\sqrt{\rho}}.$$
    By the hypothesis we have
    $$\|\Lambda^{(k)}-\lambda\|_1\leq \epsilon,$$
    and consequently for every $k$ and $j$
    $$\|\hat{\rho}^{(k)}_j-\hat{\rho}_j\|_1\leq 
                                  \frac{2}{\lambda_0^2}\epsilon.$$
    Thus we can estimate for every $k$:
    \begin{equation*}\begin{split}
      I(\Lambda^{(k)};\hat{\rho}^{(k)})
                  &=    H(\rho)-\sum_j \Lambda^{(k)}_j H(\hat{\rho}^{(k)}_j) \\
                  &\geq H(\rho)-\sum_j \lambda_j H(\hat{\rho}_j) \\
                  &\phantom{\geq\geq}-\epsilon\log d
                    +\frac{2\epsilon}{\lambda_0^2}\log\frac{2\epsilon}
{\lambda_0^2d},
    \end{split}\end{equation*}
    where we have used lemma~\ref{lemma:H:continuity}, and we are done.
  \end{beweis}

\section{Thrifty measurements}
  \label{sec:upper}
  We will prove theorem~\ref{thm:POVM:compr:1} in several steps
  (propositions~\ref{prop:one}, \ref{prop:two}, \ref{prop:three},
  and~\ref{prop:four} below).
  The strategy is as follows: we construct a series of
  sub--POVMs\footnote{A \emph{sub--POVM} is a POVM except
    for the weaker condition that the sum of its elements is only
    \emph{upper bounded} by $\1$.}
  ${\bf B}$, ${\bf C}$, ${\bf D}$, and ${\bf E}$,
  each in turn satisfying the condition
  C3 (which we demonstrate for didactical reasons even though
  this is not necessary for the ultimate proof),
  and of increasing regularity. The last step to
  construct ${\bf A}=(A_{j^l_\mu})_{\mu=1,\ldots,M}$ is a random
  selection argument with a novel large deviation probability
  estimate.
  \par
  To do this we have
  first to review the concepts of typical subspace and conditional
  typical subspace, in the form of~\cite{winter:ieee_strong}:
  \par
  For a state $\rho$ fix eigenstates $e_1,\ldots,e_d$ and
  define for $\delta>0$ the \emph{typical projector} as
  $$\Pi^l_{\rho,\delta}=\sum_{t^l\text{ with }|\sum_{k=1}^l e_{t_k} -l\rho|
                              \leq\delta\sqrt{l}\sqrt{\rho(\1-\rho)}}
                         e_{t_1}\otimes\cdots\otimes e_{t_l}.$$
  For a collection of states $\hat{\rho}_j$, $j=1,\ldots,m$, and
  $j^l\in[m]^l$ define the \emph{conditional typical projector} as
  $$\Pi^l_{\hat{\rho},\delta}(j^l)=\bigotimes_j 
\Pi^{I_j}_{\hat{\rho}_j,\delta},$$
  where $I_j=\{k:j_k=j\}$ and $\Pi^{I_j}_{\hat{\rho}_j,\delta}$
  is meant to denote the typical projector of the state $\hat{\rho}_j$
  in the positions given by $I_j$ in the tensor product of $l$ factors.
  From~\cite{winter:ieee_strong} we cite the following 
  properties of these projectors:
  \begin{align}
    \label{eq:typical:upper}
    \tr\Pi^l_{\rho,\delta}            &\leq \exp\left(lH(\rho)+Kd\delta\sqrt{l}\right), \\
    \label{eq:typical:lower}
    \tr\Pi^l_{\rho,\delta}            &\geq \left(1-\frac{d}{\delta^2}\right)
                                            \exp\left(lH(\rho)-Kd\delta\sqrt{l}\right), \\
    \label{eq:cond:typical:upper}
    \tr\Pi^l_{\hat{\rho},\delta}(j^l) &\leq \exp\left(lH(\hat{\rho}|P_{j^l})
                                                             +Kmd\delta\sqrt{l}\right), \\
    \label{eq:cond:typical:lower}
    \tr\Pi^l_{\hat{\rho},\delta}(j^l) &\geq \left(1-\frac{md}{\delta^2}\right)
                                            \exp\left(lH(\hat{\rho}|P_{j^l})
                                                             +Kmd\delta\sqrt{l}\right),
  \end{align}
  for an absolute constant $K>0$, and the empirical distribution
  $P_{j^l}$ of letters $j$ in the word $j^l$:
  $$P_{j^l}(j)=\frac{N(j|j^l)}{l}=\frac{\text{\# of occurences of }j\text{ in }j^l}{l}.$$
  Also from~\cite{winter:ieee_strong}:
  \begin{align}
    \label{eq:typical:prob}
    \tr(\rho^{\otimes l}\Pi^l_{\rho,\delta})            &\geq 1-\frac{d}{\delta^2}, \\
    \label{eq:cond:typical:prob}
    \tr(\hat{\rho}_{j^l}\Pi^l_{\hat{\rho},\delta}(j^l)) &\geq 1-\frac{md}{\delta^2},
  \end{align}
  with $r$ denoting the minimal eigenvalue of $\rho$.
  \par
  To end this review observe the following important operator
  estimates:
  \begin{align}
    \label{eq:typical:deletion}
    \Pi^l_{\rho,\delta}
             &\geq \Pi^{[l]\setminus k}_{\rho,\delta'}\otimes \1, \\
    \label{eq:cond:typical:deletion}
    \Pi^l_{\hat{\rho},\delta}(j^l)
             &\geq \Pi^{[l]\setminus k}_{\hat{\rho},\delta'}
                      (j^{[l]\setminus k})\otimes \1,
  \end{align}
  where $\delta'=\delta-1/r\geq\delta/2$, if we assume $\delta\geq 2/r$.
  Inequalities (\ref{eq:typical:deletion}) and 
  (\ref{eq:cond:typical:deletion}) will be  
  used in conjunction with the following lemma:
  \begin{lemma}
    \label{lemma:partialtr:inequ}
    Let $C$ be a positive operator on ${\cal H}_1\otimes{\cal H}_2$,
    $\Pi$ a projector on ${\cal H}_1\otimes{\cal H}_2$, and $\Pi_0$
    a projector on ${\cal H}_2$ such that $\Pi\geq\1\otimes\Pi_0$.
    Then
    $$\tr_2\left(\Pi C\Pi\right)
                 \geq \tr_2\left((\1\otimes\Pi_0)C(\1\otimes\Pi_0)\right).$$
  \end{lemma}
  \begin{proof}
    Because of $\Pi(\1\otimes\Pi_0)=\1\otimes\Pi_0$ we may assume that
    $C=\Pi C\Pi$. Thus we have to prove that
    $$\tr_2 C\geq \tr_2\left((\1\otimes\Pi_0)C(\1\otimes\Pi_0)\right).$$
    But this is equivalent to
    $$\forall A\geq 0\
    \tr\left((A\otimes\1)C\right)\geq\tr\left(A\otimes\Pi_0
      C\right),$$ 
    which in turn is equivalent to $A\otimes\1\geq A\otimes\Pi_0$,
    and this is obvious.
    \qed
  \end{proof}

  Define the following operators (with $\rho$ and $\hat{\rho}_j$ as in
  section~\ref{sec:model}):
  $$B_{j^l}=\sqrt{\rho^{-1}}^{\otimes l}\Pi^l_{\hat{\rho},\delta}(j^l)
              \sqrt{\rho}^{\otimes l}a_{j^l}\sqrt{\rho}^{\otimes l}
            \Pi^l_{\hat{\rho},\delta}(j^l)\sqrt{\rho^{-1}}^{\otimes l}.$$
  Intuitively this means to confine the $a_{j^l}$ to the range of the
  conditional typical projector $\Pi^l_{\hat{\rho},\delta}(j^l)$.
  \begin{prop}
    \label{prop:one}
    \begin{enumerate}
      \item $0\leq B_{j^l}\leq a_{j^l}$.
      \item $\tr(\rho^{\otimes l}B_{j^l})\geq 
                 \left(1-\frac{md}{\delta^2}\right)\tr(\rho^{\otimes l}a_{j^l})$.
      \item $\sqrt{\rho}a_j\sqrt{\rho}-\Delta^1_j
                \leq \sqrt{\rho}B^{(k)}_j\!\sqrt{\rho}
                   \leq \sqrt{\rho}a_j\sqrt{\rho}$,
        with $\Delta^1_j\geq 0$ and
        $\tr\Delta^1_j\leq \lambda_j\frac{md}{\delta^2}.$
      \item $\forall k\sum_j \|\sqrt{\rho}B^{(k)}_j\!\sqrt{\rho}
                                         -\sqrt{\rho}a_j\sqrt{\rho}\|_1
                                                \leq \frac{md}{\delta^2}$.
      \item $\forall k\sum_j \|B^{(k)}_j-a_j\|\leq \frac{md}{r\delta^2}$.
    \end{enumerate}
  \end{prop}
  \begin{proof}
    1. is equivalent to
    $\sqrt{\rho}^{\otimes l}B_{j^l}\sqrt{\rho}^{\otimes l}
      \leq \sqrt{\rho}^{\otimes l}a_{j^l}\sqrt{\rho}^{\otimes l}$,
    which is immediate from the definition.
    \\
    2. is essentially equation~\ref{eq:cond:typical:prob}.
    \\
    3. follows from 1. and 2.
    \\
    Finally, 4. and 5. are easy consequences of 3.
    \qed
  \end{proof}
  \par
  Defining the operators
  $$C_{j^l}=\Pi^l_{\rho,\delta}B_{j^l}\Pi^l_{\rho,\delta},$$
  i.e. restricting the $B_{j^l}$ to the range of the typical projector
  $\Pi^l_{\rho,\delta}$, we find
  \begin{prop}
    \label{prop:two}
    \begin{enumerate}
      \item $\tr(\rho^{\otimes l}C_{j^l})\leq \tr(\rho^{\otimes l}B_{j^l})$.
      \item $\sqrt{\rho}a_j\sqrt{\rho}-\Delta^2_j
                \leq \sqrt{\rho}C^{(k)}_j\!\sqrt{\rho}
                   \leq \sqrt{\rho}a_j\sqrt{\rho}+\Delta^2$,
        with $\Delta^2_j\geq 0$, $\Delta^2=\sum_j \Delta^2_j$,
        and $\tr\Delta^2_j\leq \lambda_j\frac{md+4d}{\delta^2}.$
      \item $\forall k\sum_j \|\sqrt{\rho}C^{(k)}_j\!\sqrt{\rho}-
                                        \sqrt{\rho}B^{(k)}_j\!\sqrt{\rho}\|_1
                                                      \leq 
\frac{m^2+4md}{\delta^2}$.
      \item $\forall k\sum_j \|C^{(k)}_j-B^{(k)}_j\|\leq 
\frac{m^2d+4md}{r\delta^2}$.
    \end{enumerate}
  \end{prop}
  \begin{proof}
    1. follows from
    $\Pi^l_{\rho,\delta}\rho^{\otimes l}\Pi^l_{\rho,\delta}\leq 
\rho^{\otimes l}$,
    and the definition.
    \\
    To prove 2., we first do the lower bound (the other follows from this
    straightforwardly):
    \begin{equation*}\begin{split}
      \sqrt{\rho}C^{(k)}_j\sqrt{\rho}
           &=\tr_{\neq k}\left(\sum_{j^l:\ j_k=j} \sqrt{\rho}^{\otimes l}
                                            C_{j^l}\sqrt{\rho}^{\otimes l}
\right)\\
           &=\tr_{\neq k}\!\left(\!\Pi^l_{\rho,\delta}\!
\left(\sum_{j^l:\ j_k=j} 
                          \sqrt{\rho}^{\otimes l}B_{j^l}
\sqrt{\rho}^{\otimes l}\right)
                                                      \!\Pi^l_{\rho,\delta}
\right)\\
           &\!\!\!\!\!\!\!\!\!\!\!\!\!\!\!\!\!
             \geq\tr_{\neq k}\!\left(\!\!\left(\sum_{j^l:\ j_k=j}
                  \sqrt{\rho}^{\otimes l}a_{j^l}\sqrt{\rho}^{\otimes
                    l}-
\Delta_{j^l}
                   \!\right)\!\!(\Pi^{[l]\setminus
                   k}_{\rho,\delta}\otimes\1)
\!\right)\\
           &=\sqrt{\rho}a_j\sqrt{\rho}\,\,
\tr\left(\rho^{\otimes [l]\setminus k}
                                   \Pi^{[l]\setminus
                                     k}_{\rho,\delta}\right)-
\Delta\\
           &=\sqrt{\rho}a_j\sqrt{\rho}-\Delta^2_j,
    \end{split}\end{equation*}
    where the inequality is with
    $\Delta_{j^l}\geq 0$, $\tr\Delta_{j^l}\leq\frac{md}{\delta^2}\lambda_{j^l}$
    (by proposition~\ref{prop:one}.1. and 2.), and by
    lemma~\ref{lemma:partialtr:inequ}. Hence the subsequent equalities
    are valid with
    $$\tr\Delta\leq \lambda_j\frac{md}{\delta^2},\text{ with }
                                     \Delta=\sum_{j^l:\ j_k=j} \Delta_{j^l},$$
    and
    $$\Delta^2_j=\Delta+
                 \sqrt{\rho}a_j\sqrt{\rho}
\left(1-\tr\left(\rho^{\otimes [l]\setminus k}
                                     \Pi^{[l]\setminus
                                       k}_{\rho,\delta}\right)
\right).$$
    By equation~\ref{eq:typical:prob}
    we conclude $\tr\Delta^2_j\leq\lambda_j\frac{md+4d}{\delta^2}$.
    \\
    Finally, 3. and 4. are easy consequences of 2.
    \qed
  \end{proof}
  \par
  Now with the probabilities $\lambda_j=\tr(\rho a_j)$ define the
  \emph{set of typical sequences}
  $${\cal T}^l_\delta=\{j^l:\forall j\ |N(j|j^l)-l\lambda_j|\leq
                            \delta\sqrt{l}\sqrt{\lambda_j(1-\lambda_j)}\}.$$
  \par
  The next simplification is to use only
  operators of our sub--POVM ${\bf C}$ with
  \emph{typical} $j^l$:
  define the sub--POVM ${\bf D}$ to consist of the $C_{j^l}$ for
  $j^l\in{\cal T}^l_\delta$, i.e. ${\bf D}=
(C_{j^l})_{j^l\in{\cal T}^l_\delta}$.
  \begin{prop}
    \label{prop:three}
    \begin{enumerate}
      \item $\lambda^{l}({\cal T}^l_\delta)=:S\geq 1-\frac{m}{\delta^2}$.
      \item $\tr(\rho^{\otimes l}D_{j^l})\geq 
           \left(1-\frac{2m^3 d}{r^2 \delta^2}\right)
\tr(\rho^{\otimes l}a_{j^l})$.
      \item $\sqrt{\rho}D^{(k)}_j\!\sqrt{\rho}=
\sqrt{\rho}C^{(k)}_j\!\sqrt{\rho}
                                               -\Delta^3_j$,
        with $\Delta^3_j\geq 0$, $\sum_j \tr\Delta^3_j\leq m/\delta^2$.
      \item $\forall k\sum_j \|\sqrt{\rho}D^{(k)}_j\!\sqrt{\rho}-
                                      \sqrt{\rho}C^{(k)}_j\!\sqrt{\rho}\|_1
                                                         \leq 
\frac{m}{\delta^2}$.
      \item $\forall k\sum_j \|D^{(k)}_j-C^{(k)}_j\|\leq \frac{m}{r\delta^2}$.
    \end{enumerate}
  \end{prop}
  \begin{proof}
    1. follows from Chebyshev's inequality (compare~\cite{winter:ieee_strong}).
    \\
    2. is seen as follows: with the eigenstates $e_t$ of $\rho$
    define
    $$\tilde{\rho}_j=E(\hat{\rho}_j)=\sum_t e_t\hat{\rho}_j e_t,$$
    with the conditional expectation $E$. Then it is obvious that
    $$\tr\left(\hat{\rho}_{j^l}\Pi^l_{\rho,\delta}\right)
         =\tr\left(\tilde{\rho}_{j^l}\Pi^l_{\rho,\delta}\right).$$
    From the definitions it can be directly verified that,
    with $\tilde{\rho}=\frac{1}{l}\sum_k \tilde{\rho}_{j_k}$,
    $$\Pi^l_{\rho,\delta}\geq\Pi^l_{\tilde{\rho},\frac{r}{m}\delta\sqrt{m}},$$
    hence by~\cite{winter:ieee_strong}, lemma V.9
    $$\tr\left(\hat{\rho}_{j^l}\Pi^l_{\rho,\delta}\right)
                                           \geq 1-\frac{m^3 d}{r^2 \delta^2},$$
    and with proposition~\ref{prop:one}.2. the claim follows.
    \\
    For 3. observe
    $$\sqrt{\rho}C^{(k)}_j\!\sqrt{\rho}-\sqrt{\rho}D^{(k)}_j\!\sqrt{\rho}
       =\!\!\sum_{j^l\not\in{\cal T}^l_\delta:\ j_k=j}\!\!
               \tr_{\neq k} \sqrt{\rho}^{\otimes l}C_{j^l}
                             \sqrt{\rho}^{\otimes l},$$
    and denoting the r.h.s by $\Delta^3_j$, the claim follows from 1.,
    observing that
    $$\tr\sqrt{\rho}^{\otimes l}C_{j^l}\sqrt{\rho}^{\otimes l}
             \leq \lambda_{j^l},$$
    by propositions~\ref{prop:one}.1. and~\ref{prop:two}.1.
    \\
    Again, 4. and 5. are easy consequences.
    \qed
  \end{proof}
  \par
  We shall use the probability distribution $\Lambda$ on
  ${\cal T}^l_\delta$, with
  $$\Lambda_{j^l}=\frac{1}{S}\lambda^l_{j^j}.$$
  Observe that
  $$\omega=\sum_{j^l\in{\cal T}^l_\delta} \sqrt{\rho}^{\otimes l}
                         D_{j^l}\sqrt{\rho}^{\otimes l}
            =\rho^{\otimes l}-\Delta^4,$$
  with
  $$\tr\Delta^4\leq \frac{(m+1)(d+1)}{\delta^2}=:c.$$
  Introducing
  $$\alpha=\left(1-\frac{d}{\delta^2}\right)\exp\left(-lH(\rho)-
                  Kd\delta\sqrt{l}\right)$$
  (so that
  $\Pi^l_{\rho,\delta}\rho^{\otimes l}\Pi^l_{\rho,\delta}
      \geq\alpha\Pi^l_{\rho,\delta}$),
  we can construct the subspace spanned by the eigenvectors of $M$
  corresponding to eigenvalues at least $c\alpha$. With its projection $\Pi$
  we have $\Pi\omega\Pi\geq c\alpha\Pi$. This implies
  $$\tr\left(\omega(\Pi^l_{\rho,\delta}-\Pi)\right)\leq c,$$
  hence because of $\tr\omega\Pi^l_{\rho,\delta}\geq 1-c$
  \begin{equation}
    \label{eq:pi:weight}
    \tr\omega\Pi\geq 1-2c.
  \end{equation}
  \par
  Now, define the sub--POVM ${\bf E}$ by
  $$E_{j^l}=\sqrt{\rho^{-1}}^{\otimes l}\Pi\sqrt{\rho}^{\otimes l}
                                D_{j^l}
            \sqrt{\rho}^{\otimes l}\Pi\sqrt{\rho^{-1}}^{\otimes l},$$
  for $j^l\in{\cal T}^l_\delta$.
  \begin{prop}
    \label{prop:four}
    For $j^l\in{\cal T}^l_\delta$:
    \begin{enumerate}
      \item $\tr(\rho^{\otimes l}E_{j^l})\leq \tr(\rho^{\otimes l}D_{j^l})$.
      \item $\forall k\sum_j \|\sqrt{\rho}E^{(k)}_j\!\sqrt{\rho}-
                                      \sqrt{\rho}D^{(k)}_j\!\sqrt{\rho}\|_1
                                                         \leq 2mc$.
      \item $\forall k\sum_j \|E^{(k)}_j-D^{(k)}_j\|_1\leq 2mc/r$.
    \end{enumerate}
  \end{prop}
  \begin{proof}
    1. is obvious, and 3. follows from 2.
    \\
    To prove 2., first calculate
    \begin{equation*}\begin{split}
      \sqrt{\rho}E^{(k)}_j\sqrt{\rho}
              &= \tr_{\neq k}\left(\sum_{j^l\in{\cal T}^l_\delta:\ j_k=j}
                     \sqrt{\rho}^{\otimes
                       l}E_{j^l}\sqrt{\rho}^{\otimes l}\right)\\
              &= \tr_{\neq k}\left(\sum_{j^l\in{\cal T}^l_\delta:\ j_k=j}
                     \Pi\sqrt{\rho}^{\otimes l}D_{j^l}
                        \sqrt{\rho}^{\otimes l}\Pi\right)\\
              &=:\lambda_j\tr_{\neq k}\Pi\omega_{kj}\Pi,
    \end{split}\end{equation*}
    with
    $$\omega_{kj}=\frac{1}{\lambda_j}\sum_{j^l\in{\cal T}^l_\delta:\ j_k=j}
                     \sqrt{\rho}^{\otimes l}D_{j^l}\sqrt{\rho}^{\otimes l}.$$
    Observe that by equation~\ref{eq:pi:weight}
    $$\tr\omega_{kj}\Pi\geq 1-\frac{2c}{\lambda_j}.$$
    But because of
    \begin{equation*}\begin{split}
      \Pi^l_{\rho,\delta}\omega_{kl}\Pi^l_{\rho,\delta}-\Pi\omega_{kl}\Pi
             &=(\Pi^l_{\rho,\delta}-\Pi)\omega_{kl}
                                        \Pi^l_{\rho,\delta}            \\
             &\phantom{=}+\Pi\omega_{kl}(\Pi^l_{\rho,\delta}-\Pi),
    \end{split}\end{equation*}
    we get
    \begin{equation*}\begin{split}
      \|\Pi^l_{\rho,\delta}\omega_{kl}\Pi^l_{\rho,\delta}-\Pi\omega_{kl}\Pi\|_1
           &\leq\|(\Pi^l_{\rho,\delta}-\Pi)\omega_{kl}
                              \Pi^l_{\rho,\delta}\|_1     \\
           &\phantom{=}+\|\Pi\omega_{kl}
                (\Pi^l_{\rho,\delta}-\Pi)\|_1             \\
           &\leq 2\tr\left(\omega_{kl}
              (\Pi^l_{\rho,\delta}-\Pi)\right)            \\
           &\leq 2c/\lambda_j,
    \end{split}\end{equation*}
    thus we conclude
    \begin{equation*}\begin{split}
      \sqrt{\rho}E^{(k)}_j\sqrt{\rho}
              &=\tr_{\neq k}\Pi^l_{\rho,\delta}\lambda_j\omega_{kj}
                             \Pi^l_{\rho,\delta}+\Delta^5_j  \\
              &=\sqrt{\rho}D^{(k)}_j\sqrt{\rho}+\Delta^5_j,
    \end{split}\end{equation*}
    where $\|\Delta^5_j\|_1\leq 2c.$
    \qed
  \end{proof}
  \par
  The proof of the theorem will now be completed by a random selection
  of a sufficient number of elements from ${\bf E}$.
  We invoke a result from~\cite{winter:operatorRV}:
  \begin{lemma}
    \label{lemma:op:large:deviation}
    Let $X_1,\ldots,X_M$ be i.i.d. random variables with
    values in the algebra ${\cal L}({\cal K})$ of linear operators
    on ${\cal K}$, which are bounded between
    $0$ and $\1$. Assume that the average $\E X_\mu=\sigma\geq s\1$.
    Then for every $\eta>0$
    $$\Pr\left\{\frac{1}{M}\sum_{\mu=1}^M X_\mu\not\leq(1+\eta)\sigma\right\}
          \leq \dim{\cal K}\exp\left(-M\frac{\eta^2 s}{2\ln 2}\right)\!.$$
  \end{lemma}
  \par
  With this we can now finish \\
  \begin{beweis}{of theorem~\ref{thm:POVM:compr:1}}
    Starting from the POVM ${\bf a}^{\otimes l}$ construct the
    sub--POVMs ${\bf B}$, ${\bf C}$, ${\bf D}$, and ${\bf E}$, as above.
    \par
    Define i.i.d. random variables $J_1,\ldots,J_M$ with
    values in ${\cal T}^l_\delta$ such that
    $$\Pr\{J_\mu=j^l\}=\Lambda_{j^l},\quad \mu=1,\ldots,M.$$
    These define operator valued random variables
    $$X_\mu=\frac{S}{\lambda_{J_\mu}}
                \sqrt{\rho}^{\otimes l}E_{J_\mu}\sqrt{\rho}^{\otimes l},
                                                         \quad 
                      \mu=1,\ldots,M.$$
    Observe that for all $\mu$
    $$\E X_\mu=\Pi\omega\Pi\leq\omega
                           \leq\Pi^l_{\rho,\delta}\rho^{\otimes l}
                                     \Pi^l_{\rho,\delta}
                           \leq\rho^{\otimes l},$$
    and that $X_\mu\geq 0$ with $\tr X_\mu\leq 1$. Furthermore, since
    $$\sqrt{\rho}^{\otimes l}E_{j^l}\sqrt{\rho}^{\otimes l}
            =\lambda_{j^l}\Pi\Pi^l_{\hat{\rho},\delta}(j^l)
                        \hat{\rho}_{j^l}\Pi^l_{\hat{\rho},\delta}(j^l)\Pi,$$
    we have $X_\mu\leq\beta\Pi$ with
    $$\beta=\exp\left(-lH(\hat{\rho}|\lambda)+Kmd\delta\sqrt{l}\right).$$
    Most importantly we find
    $$\E X_\mu=\Pi\omega\Pi\geq c\alpha\Pi.$$
    Apply lemma~\ref{lemma:op:large:deviation} to the variables
    $\beta^{-1}X_\mu$ to find
    \begin{equation}\begin{split}
      \label{eq:prob:one}
      \Pr&\left\{\frac{1}{M}\sum_{\mu=1}^M X_\mu
                              \not\leq (1+\eta)\Pi\omega\Pi\right\}          \\
         &\phantom{============}\leq
                  \tr\Pi\exp\left(-M\frac{\eta^2 c\alpha}{2\beta\ln 2}\right).
    \end{split}\end{equation}
    \par
    Define $Y^{[jk]}_{\mu}=\tr_{\neq k}X_\mu$
    if $J_{\mu k}=j$ and $0$ otherwise. Observe that
    $$\E Y^{[jk]}_{\mu}=\sqrt{\rho}E^{(k)}_j\sqrt{\rho},$$
    so by propositions~\ref{prop:one}.2., \ref{prop:two}.2.,
    \ref{prop:three}.2., and~\ref{prop:four}.2.
    $$\|\E Y^{[jk]}_{\mu}-\sqrt{\rho}a_j\sqrt{\rho}\|_1\leq
      2c+\frac{m}{\delta^2}+\frac{m^2 d+4md}{\delta^2}+\frac{md}
             {\delta^2}=:\tilde{c}.$$
    Thus, by the operator Chebyshev inequality~\cite{winter:operatorRV}
    \begin{equation}
      \label{eq:prob:two}
      \Pr\left\{\left\|\sum_\mu Y^{[jk]}_\mu-M\sqrt{\rho}a_j\sqrt{\rho}\right\|_1
        \!\! >M\tilde{c}+\delta\sqrt{M}\sqrt{l}\right\}\leq \frac{d}{l\delta^2}.
    \end{equation}
    In case that the sum of the right hand sides of the probability
    estimates from equations (\ref{eq:prob:one}) and (\ref{eq:prob:two})
    ($j=1,\ldots,m$, $k=1,\ldots,l$) is less than $1$
    --- which can be forced by choosing
    \begin{equation}
      \label{eq:delta:M}
      \delta>\sqrt{2md}\text{ and }
      M>\frac{2\ln 2(1-\log\alpha)}{\eta^2 c}\frac{\beta}{\alpha}
    \end{equation}
    --- there are actual values $J_1=j^l_1,\ldots,J_M=j^l_M$ such that
    \begin{equation}
      \label{eq:condition:1}
      \frac{1}{M}\sum_\mu \frac{1}{\Lambda_{j^l_\mu}}
                      \sqrt{\rho}^{\otimes l}E_{j^l_\mu}\sqrt{\rho}^{\otimes l}
       \leq (1+\eta)\Pi^l_{\rho,\delta}\rho^{\otimes l}\Pi^l_{\rho,\delta},
    \end{equation}
    and
    \begin{equation}
      \label{eq:condition:2}
      \begin{split}
      \forall k\forall j\,
         \left\|\frac{1}{M}\!\sum_{\mu:\ j_{\mu k}=j} 
              \!\!\!\tr_{\neq k}\!\left(
                \frac{1}{\Lambda_{j^l_\mu}}
                \sqrt{\rho}^{\otimes l}E_{j^l_\mu}\sqrt{\rho}^{\otimes l}
                \right)\!-\!\sqrt{\rho}a_j\sqrt{\rho}\right\|_1 & \\
                         &\hspace{-1.9cm}
                          \leq \tilde{c}+\frac{\delta\sqrt{l}}{\sqrt{M}}.
      \end{split}
    \end{equation}
    In this case we may form the following sub--POVM:
    \begin{equation*}\begin{split}
      \tilde{A}_{j^l_\mu}
        &= \frac{1}{(1+\eta)M}
             \sqrt{\rho^{-1}}^{\otimes l}\!
               \left(\frac{S}{\lambda_{j^l_\mu}}
                 \sqrt{\rho}^{\otimes l}E_{j^l_\mu}
                  \sqrt{\rho}^{\otimes l}\right)
             \!\sqrt{\rho^{-1}}^{\otimes l} \\
        &= \frac{1}{(1+\eta)M}\frac{S}{\lambda_{j^l_\mu}}E_{j^l_\mu}.
    \end{split}\end{equation*}
    First observe that this is indeed a sub--POVM, as
    by equation (\ref{eq:condition:1})
    $$\sum_\mu \tilde{A}_{j^l_\mu}\leq \Pi^l_{\rho,\delta}.$$
    We claim that it satisfies condition C3, more precisely,
    by equation (\ref{eq:condition:2}) we find
    \begin{equation*}\begin{split}
      \forall k\forall j\,
        \left\|\sum_{\mu:\ j_{\mu k}=j} \!\!\!\tr_{\neq k}\!\left(
            \sqrt{\rho}^{\otimes l}\tilde{A}_{j^l_\mu}\sqrt{\rho}^{\otimes l}
            \right)\!-\!\sqrt{\rho}a_j\sqrt{\rho}\right\|_1 & \\
                            &\hspace{-1.9cm}
                             \leq\eta+\tilde{c}+
                             \frac{\delta\sqrt{l}}{\sqrt{M}}.
    \end{split}\end{equation*}
    \par
    Distributing the remainder $R=\1-\sum_\mu \tilde{A}_{j^l_\mu}$
    equally over the operators will give us our desired
    POVM ${\bf A}$:
    $$A_{j^l_\mu}=\tilde{A}_{j^l_\mu}+\frac{1}{M}R.$$
    Namely, it is immediate that now 
    (inserting equation~\ref{eq:POVM:marginals})
    \begin{equation*}
      \forall k\forall j\
      \left\|\sqrt{\rho}A^{(k)}_j\!\sqrt{\rho}\!-
                \!\sqrt{\rho}a_j\sqrt{\rho}\right\|_1
               \!\leq (m+1)\!\left(\!\eta+\tilde{c}+
                \frac{\delta\sqrt{l}}{\sqrt{M}}\right)\!,
    \end{equation*}
  and we are done, by choosing $\eta=\delta^{-2}$, with
  $\delta$ suitably large, and $M$ according to equation (\ref{eq:delta:M}).
  \end{beweis}

\section{Extensions}
  \label{sec:extensions}
  Not to encumber the proofs with too many estimates which actually
  would not contribute to the understanding of the results, we chose
  to present theorems~\ref{thm:POVM:compr:1} and~\ref{thm:POVM:compr:lower}
  in their above form.
  \par
  However, let us see here how far we can actually
  get with our theorem~\ref{thm:POVM:compr:1}:
  we might for example go beyond C3 by requiring
  \begin{equation}
    \sum_k\sum_j \|A^{(k)}_j-a_j\|\leq\epsilon.\tag{C4}
  \end{equation}
  We might go even further, and demand that ${\bf A}$ approximates
  ${\bf a}^{\otimes l}$ not only on single factors but also on
  all subsets of factors, $K\subset\{1,\ldots,l\}$, of
  moderate growing size, say $|K|\leq\nu_l=o(l)$:
  \begin{equation}
    \sum_{K\subset[l],|K|\leq\nu_l}\sum_{j^K} 
           \|A^{(K)}_{j^K}-a_{j^K}\|\leq\epsilon.\tag{C5}
  \end{equation}
  Here
  $$A^{(K)}_{j^K}=\tr_{[l]\setminus K}\!
                 \left(\!\left(\rho^{\otimes[l]\setminus K}
                  \otimes\1^{\otimes K}\right)
                   \!\!\sum_{\mu:\ \forall k\in K\ j_{\mu k}=
         j_k}\!\! A_{j^l_\mu}\right)$$
  is the restriction of ${\bf A}$ to the tensor factors $K$ and
  $$a_{j^K}=\bigotimes_{k\in K} a_{j_k}$$
  is an element of the $K$--factor POVM ${\bf a}^{\otimes K}$.
  \par
  It turns out that, using slightly stronger estimates for
  the typical subspaces and typical sequences than those
  used in section~\ref{sec:upper} one can prove 
  \begin{thm}
    \label{thm:POVM:compr:2}
    With the above notation, there exists a POVM
    ${\bf A}=(A_{j^l_\mu})_{\mu=1,\ldots,M}$ with
    $$M\leq \exp\left(l\left(H(\rho)-\sum_j \lambda_j H(\hat{\rho}_j)
                       \right)+o(l)\right)$$
    and satisfying condition C5.
  \end{thm}
  \begin{proof}
    From~\cite{winter:diss}, lemma I.9, we use the following estimates:
    for fixed $\hat{\rho}_j$ and $\rho$ there exists a constant $\gamma>0$
    such that
    \begin{align}
      \label{eq:typical:prob:exp}
      \tr\left(\rho^{\otimes l}\Pi^l_{\rho,\delta}\right)
                     &\geq 1-d\!\cdot\! e^{-\gamma\delta^2},\\
      \label{eq:cond:typical:prob:exp}
      \tr\left(\hat{\rho}_{j^l}\Pi^l_{\hat{\rho},\delta}(j^l)\right)
                     &\geq 1-md\!\cdot\! e^{-\gamma\delta^2}.
    \end{align}
    Instead of equations (\ref{eq:typical:prob}) and (\ref{eq:cond:typical:prob}),
    use equations (\ref{eq:typical:prob:exp}) and (\ref{eq:cond:typical:prob:exp})
    in all steps of the proof in section~\ref{sec:upper}. Then, choosing
    $\delta=\delta_l$ such that
    $$\nu_l=o(\delta_l^2),\quad \delta_l=o(\sqrt{l}),$$
    and with $\eta=\exp(-\delta^2)$, the theorem follows.
    \qed
  \end{proof}
  \par
  As an immediate corollary we get an improvement of
  theorem~\ref{thm:massar:popescu}:
  \begin{thm}
    \label{thm:mp:improved}
    For $\epsilon>0$ and $l$ large enough there exists a POVM ${\bf A}$
    satisfying
    $$l\!\cdot\! F({\bf A})\geq l\!\cdot\! F_{\rm opt}-\epsilon$$
    and with
    $$M\leq\exp(l(H(\rho)+\epsilon))$$
    many outcomes.
    \qed
  \end{thm}
  \par
  On the other hand, inspection of the proof of 
  theorem~\ref{thm:POVM:compr:lower}
  shows that it remains valid (up to another 
  $O(\epsilon l)$ in the exponent)
  under the slightly weaker condition
  \begin{equation}
    \frac{1}{l}\sum_{k=1}^l \sum_j \|A^{(k)}_j-a_j\|
    \leq\epsilon.\tag{C2$\frac{1}{2}$}
  \end{equation}

\section{Discussion}
\label{sec:conclusion}
In this article we have shown how to compress quantum
measurements. More precisely we have shown how to devise a measurement 
${\bf A}$
that is close to a certain given POVM ${\bf a}$ but that produces a
minimum amount of data. This minimum amount of data is equal to 
$I(\lambda;\hat{\rho}) =
  H(\rho) - \sum_j \lambda_j H(\hat \rho_j)$,
where
$$\hat{\rho}_j = \frac{1}{\lambda_j}\sqrt{\rho} a_j\sqrt{\rho},
                                \quad \lambda_j = \tr \rho a_j.$$
\par
This result provides a precise measure of how much knowledge about
the unknown states is provided by the measurement ${\bf a}$. Namely the amount
of knowledge provided by the measurement is equal to the minimal amount of
classical data produced measurements ${\bf A}$ that are close to
${\bf a}$. This is because
the measurement ${\bf A}$ resembles the measurement  ${\bf a}$, 
hence provides as much knowledge about the states as  ${\bf a}$.
But the spurious randomness in data produced by 
the measurement  ${\bf a}$ has been removed.
Thus we deduce that the amount of
meaningful data produced by the measurement ${\bf a}$ is
$I(\lambda;\hat \rho)$. 
\par
We now consider several  questions and lines of
inquiry which are suggested by the present results.

\subsection{Information missed by incomplete measurements}
Consider a POVM that is not maximally refined. By this we mean that 
the POVM elements 
$a_j$ are not all proportional to one dimensional projectors. Such a
POVM does not provide maximum knowledge about a quantum state. However 
at a later stage one can refine the POVM so as to obtain additional
knowledge about the state. We would like to know whether carrying out
such a sequence of measurements provides the maximum knowledge about
the state, or whether their is an irreversible loss of knowledge in
such a two step measurement.
We shall argue, using the results presented in this paper, that if the 
first measurement is carried out in such a way to minimize disturbance 
to the state, then no knowledge is lost by such a two step procedure.
\par
When the POVM is not maximally refined the amount of meaningful data
produced by the measurement is 
less than if the measurement is maximally refined since the term which 
one subtracts
$\sum_j \lambda_j H(\hat{\rho}_j)$ in $I(\lambda; \hat{\rho})$
vanishes in one case and not in the other. Suppose that
after the first measurement one carries out a second measurement ${\bf b}$ 
which is maximally refined. Let us note that after the first
measurement, if the state was $\rho_i$ and the outcome was $j$, one
obtains the state $\sigma_{ji}$ given by the completely positive map
\begin{equation}
  \sigma_{ji}=\frac{\sum_\nu V_{j\nu} \rho_i V_{j\nu}^\dagger }{\tr \rho_i a_j}
  \label{CPmap1}
\end{equation}
where $\sum_\nu V_{j\nu}^\dagger V_{j\nu}= a_j$. And the average state
if the outcome was $j$ is
\begin{equation}
  \sigma_{j}= \frac{\sum_\nu V_{j\nu} \rho V_{j\nu}^\dagger}{\tr \rho a_j}
  \label{CPmap2}
\end{equation}
Since the second measurement is maximally refined, the
amount of meaningful data 
it produces is equal to $H(\sigma_{j})$.
\par
We first consider the case when the completely
positive map has only one term in its Kraus representation.
In this case
$\sigma_{j} = U_j \sqrt{a_j} \rho \sqrt{a_j}U_j^{\dagger}$
(where $U_j$ is a unitary matrix).
We now show that the amount of meaningful data produced by the second
measurement $H(\sigma_{j})$ is equal to the 
amount of data missing from the first
measurement $H(\hat\rho_j)$. This follows from the fact that 
$\sigma_j$ and $\hat{\rho}_j$
have the same spectrum, as they are conjugates. To show 
this it is sufficient to
show that $\sqrt{\rho}a_j\sqrt{\rho}$ and $\sqrt{a_j}\rho\sqrt{a_j}$ are
conjugates. Using the notation $B=\sqrt{a_j}\sqrt{\rho}$, we have
\begin{align*}
  B^\dagger B &=\sqrt{\rho}a_j\sqrt{\rho}\ , \\
  B B^\dagger &=\sqrt{a_j}\rho\sqrt{a_j}\ .
\end{align*}
Introducing the polar decomposition $B=U |B|$ (where
$U$ is unitary and $|B|=\sqrt{B^\dagger B}$), we find
$$B B^\dagger=U |B|^2 U^\dagger=U(B^\dagger B)U^\dagger$$
which is what we needed to show.
\par
Thus in the case where the Kraus representation of the completely
positive map (\ref{CPmap1}) contains only one term, the deficit in the 
amount of meaningful data produced by the first measurement 
is exactly equal to the amount of meaningful data obtained by
the second (maximally refined) measurement. It thus appears that
making an incomplete measurement which is such that the Kraus
representation of the measurement operation contains only one term
for each POVM element does not give rise to an irreversible loss of
knowledge. Rather the knowledge is still present and  can be
accessed by a second more refined measurement.
\par
The case when the Kraus representation of the completely positive
map (\ref{CPmap1})
contains only one term corresponds to the situation in which one
disturbs as little as possible the quantum state. On the other hand
when the Kraus representation contains more than one term, one easily
checks on examples that the amount of information obtained by the
second measurement bears no relation to the amount of information
obtained by the first measurement. This is because the map can
either add noise to the state or take away information.
\par
The above discussion raises an interesting question concerning the amount of
information transferred to a state or taken away from the state by a
completely positive map.
The approach developed in this paper may illuminate this
question and we hope to report on this in a future paper.

\subsection{Relation to Holevo's bound}
\label{subsec:holevo}
Consider an ensemble of states  $\{\sigma_i,\mu_i\}$
 whose average is
$\sum_i \mu_i\sigma_i = \rho$ and consider a POVM with
elements $a_j$.
We define random variables $X,Y$ with joint distribution
\begin{equation} 
  \Pr\{X=i,Y=j\}=\mu_i\tr(\sigma_i a_j) \ .
  \label{I2'}
\end{equation}
They describe the joint probability that state $\sigma_i$ occurred in
the ensemble and measurement outcome $j$ occurred.
\par
Holevo's bound~\cite{holevo:bound}
states that the mutual entropy between a source with
ensemble $\{\sigma_i,\mu_i\}$ and a measurement $a_j$ is bounded by
the entropy defect of the ensemble:
\begin{equation} 
  I(X\wedge Y)\leq I(\mu ; \sigma)
                 = H(\rho) - \sum_i \mu_i H(\sigma_i).
  \label{Hbound}
\end{equation}
\par
Note that Holevo's bound is a function only of
the ensemble $\{\sigma_i,\mu_i\}$, and the measurement plays no 
role in the bound. On the other hand in the present
paper the ensemble plays a secondary role, and we have considered
how the joint distribution of $X,Y$ changes when one changes the measurement.
In order to make connection with Holevo's bound we shall use a trick
that allows us to switch the role of ensemble and measurement.
\par
Let us denote the triple consisting of the states $\sigma_i$, the
probabilities $\mu_i$ and the POVM elements $a_j$ by
$$M_{\sigma,\mu,{\bf a}}=\{ \sigma_i , \mu_i , a_j \}.$$
We now construct a second triple
$$N_{\hat{\rho},\lambda,{\bf S}}= \{ \hat \rho_j , \lambda_j , S_i \}$$
canonically associated with the first. In this second triple the
states are
$$\hat{\rho}_j = \frac{1}{\lambda_j}\sqrt{\rho} a_j \sqrt{\rho}$$
and their probabilities are 
$\lambda_j = \tr \rho a_j$.
The POVM elements $S_i$ 
of the second triple are  the ``pretty good'' measurement of the ensemble
$\{\sigma_i,\mu_i\}$:
$$S_i=\sqrt{\rho^{-1}}\mu_i\sigma_i\sqrt{\rho^{-1}}.$$
\par
We call these two triples canonically associated for two
reasons. First the average of the states is the same
$\sum_i \mu_i \sigma_i = \sum_j \lambda_j \hat \rho_j = \rho$. Second
the probability that state $\sigma_i$ occurred 
 and measurement outcome $j$ occurred the first triple is equal to the
probability that state $\hat \rho_j$ occurred  and
the measurement outcome $i$ occurred in the second triple:
\begin{equation} 
  \Pr\{X=i,Y=j\}=\mu_i\tr(\sigma_i a_j) = \lambda_j\tr(S_i \hat \rho_j).
  \label{I2}
\end{equation}
\par
Using the relation between these two triples and in particular eq. 
(\ref{I2}) we can write two forms of Holevo's bound. The first is
equation (\ref{Hbound}), the second is 
\begin{equation} 
  I(X\wedge Y)\leq I(\hat{\rho};\lambda)
                 = H(\rho) - \sum_j \lambda_j H(\hat \rho_j).
  \label{Hbound2}
\end{equation}
Thus for a given triple, say $M_{\sigma,\mu,{\bf a}}$, we can derive two
bounds on the mutual information, the first (\ref{Hbound}) depends only 
the ensemble $\{\sigma_i , \mu_i\}$, the second (\ref{Hbound2}) depends
only on the average state $\rho$ and the POVM $\bf a$.
\par
In order to establish a connections between the present work and
Holevo's bound, we use the second form of Holevo's bound, 
equation (\ref{Hbound2}). 
Let us first 
note that theorem~\ref{thm:POVM:compr:1} shows that
one can
always devise a measurement ${\bf A}$ acting collectively on many
independent states whose marginals are close to the POVM $\bf a$
and with a
number of outcomes  equal to the right hand side of (\ref{I2}). Thus
Holevo's bound and theorem~\ref{thm:POVM:compr:1} are consistent.
\par
However we can go further and
use (\ref{I2}) together with
theorem~\ref{thm:POVM:compr:1} 
to derive Holevo's bound.
Let $\{\hat{\rho}_j,\lambda_j\}$ be any ensemble of states with average
$\rho=\sum_j \lambda_j \hat{\rho}_j$, and $(S_i)$ a POVM. With
random variables $X,Y$ as defined in the second equality in 
(\ref{I2'}), Holevo's bound is equivalent to
(\ref{Hbound2}).
Let us denote the classical mutual information $I(X\wedge Y)$
by $I(\{\hat{\rho}_j,\lambda_j\}\wedge (S_i))$.
Now we revert the argument from the beginning of
this subsection and invent the POVM ${\bf a}$ and the ensemble
$\{\sigma_i,\mu_i\}$, so that the first equality in (\ref{I2}) is satisfied.
In particular we get
$$I(\{\sigma_i,\mu_i\}\wedge 
(a_j))=I(\{\hat{\rho}_j,\lambda_j\}\wedge (S_i)),$$
and what we have to prove transforms into
$$I(\{\sigma_i,\mu_i\}\wedge (a_j))\leq I(\lambda;\hat{\rho}).$$
\par
Here our theorem~\ref{thm:POVM:compr:1}
comes in: define, for any POVM ${\bf a}$, the
fidelity function
$$F({\bf a})=I(\{\sigma_i,\mu_i\}\wedge (a_j)),$$
and for the POVM ${\bf A}$ on ${\cal H}^{\otimes l}$ the fidelity
on blocks
$$F({\bf A})=\frac{1}{l}\sum_{k=1}^l I(\{\sigma_i,\mu_i\}\wedge (A^{(k)}_j)).$$
Observe that this is a nonlinear continuous function of the POVM (the
ensemble $\{\sigma_i,\mu_i\}$ we now consider as fixed).
\par
By theorem~\ref{thm:POVM:compr:1} we find, for $\epsilon>0$ and
large enough $l$:
\begin{equation*}\begin{split}
  I(\lambda;\hat{\rho})+\epsilon
            &\geq \frac{1}{l}\log M                      \\
            &\geq \frac{1}{l}I(\{\sigma_i,\mu_i\}^{\otimes l} 
\wedge A_{j^l_\mu}) \\
            &=    \frac{1}{l}I(X^l\wedge Y)              \\
            &\phantom{===}
             \left[\Pr\{X^l=i^l,Y=\mu\}=\mu_{i^l}
\tr(\sigma_{i^l}A_{j^l_\mu})\right] \\
            &\geq \frac{1}{l}\sum_{k=1}^l I(X_k\wedge Y) \\
            &\geq \frac{1}{l}\sum_{k=1}^l I(X_k\wedge f_k(Y))
               \phantom{===.}
               \Big[f_k(\mu)=j_{\mu k}\Big]              \\
            &=    \frac{1}{l}\sum_{k=1}^l I(\{\sigma_i,\mu_i\}
\wedge (A^{(k)}_j)) \\
            &=    F({\bf A})                             \\
            &\geq F({\bf a})-\epsilon                    \\
            &=    I(\{\sigma_i,\mu_i\}\wedge (a_j))-\epsilon.
\end{split}\end{equation*}
(Only classical information inequalities have been used: the second line is by
data processing, the fourth from independence of the $X_k$, the fifth
by data processing again).
Because $\epsilon>0$ was arbitrary, we are done.

\subsection{Data vs.~information}
The above discussion concerning 
the Holevo bound can be used to address the relation
between \emph{data} and \emph{(mutual) information}.
\par
Holevo's bound as usually presented is a function only of the ensemble 
$\{\sigma_i , \mu_i \}$
of states emitted by a source. Maximizing over the measurement, with
fixed ensemble, yields the accessible information at fixed ensemble
$I_{\rm acc} (\mu;\sigma)$. It was
shown in~\cite{holevo:bound} that the accessible information attains
the Holevo bound if and only if all the states that compose the
ensemble commute. Furthermore this difference remains even
asymptotically when one considers measurements
on many independent states emitted by the source
because (see~\cite{holevo:bound:1})
$$I_{\rm acc} (\mu^{\otimes l};\sigma^{\otimes l})
   = l\cdot I_{\rm acc} (\mu;\sigma).$$
On the other hand it is known that one can carry out block coding and
construct an ensemble whose marginals are such that they are
distributed in the same way as the original ensemble, such that for
this ensemble the accessible information approaches with the Holevo bound:
see~\cite{holevo:coding,schumacher:westmoreland,winter:ieee_strong}.
\par
Let us now transcribe these results in terms of measurements, using
the second form of Holevo's bound discussed above.
If one keeps the measurement $\bf a$ fixed and maximizes over the ensemble
(with the average state $\rho$ fixed), 
one reaches the accessible information at fixed measurement and fixed
average state, which we denote $J_\rho ({\bf a})$.
It follows from the above discussion that $J_\rho ({\bf a})$ is
strictly less than $I(\lambda; \hat{\rho})$ except if all the
$\hat{\rho}_j$ commute, and that this gap remains even asymptotically since
$$J_{\rho^{\otimes l}}({\bf a}^{\otimes l})=l\cdot J_\rho({\bf a}).$$
Thus the mutual information at fixed measurement and fixed average
state is in general strictly less than the amount of meaningful 
data produced by
the measurement.
\par
However, it follows from the results
of~\cite{holevo:coding,schumacher:westmoreland,winter:ieee_strong} that 
there exists a measurement ${\bf \tilde A}$ acting on the tensor product
${\cal H}^{\otimes l}$ of the Hilbert 
space of the composite ensemble, such that its marginals are very close
to ${\bf a}$, and such that the accessible information 
$J_{\rho^{\otimes l}}({\bf \tilde A})$ equals
$l\cdot I(\lambda;\hat{\rho})$ asymptotically.
\par
We conjecture that the POVM ${\bf A}$ constructed in 
theorem~\ref{thm:POVM:compr:1} 
has all the properties of ${\bf \tilde  A}$
enumerated above. This would mean that the compressed version ${\bf A}$ 
of the POVM ${\bf a}^{\otimes l}$ asymptotically closes the gap between
mutual information and amount of data.

\subsection{Open questions}
There remain a number of open questions for future research
of which we point out a few. The first three concern a better
understanding of the conditions under which we get our result:
  \begin{enumerate}
    \item In the case where the ensemble on which the measurement is
      carried out is composed of mixed states, can one decrease
      further the amount of data produced by the measurement? The
      results proven in this paper use condition C3 in which only the
      average density matrix $\rho$ of the states enters (through the
      definition of the marginal POVMs). However it is possible, if one uses
      the weaker conditions C0, C1, or C2 that the measurements can be 
      further compressed.

    \item Conversely, one could prove that further compression is
      impossible (theorem~\ref{thm:POVM:compr:lower}) using 
      conditions C0, C1 or C2.

    \item In the case of rank--one POVM the entropy defect in
      theorems~\ref{thm:POVM:compr:1} and \ref{thm:POVM:compr:lower}
      becomes the entropy of $\rho$, the number of outcomes
      of the compressed measurement is comparable to the dimension
      of the typical subspace of $\rho^{\otimes l}$. Since the interesting
      part of the construction is in the typical subspace we may ask
      whether one can achieve the bound of
      theorem~\ref{thm:POVM:compr:1} (or a slightly weaker one)
      by a \emph{von Neumann} measurement. The methods used in the
      present paper and in~\cite{massar:popescu} do not seem to
      yield this.
\end{enumerate}

A final question concerns the 
tradeoff between fidelity and number of outcomes
of ${\bf A}$. Here we studied only the extremal case where the
fidelity should be arbitrarily close to the maximum, but
comparison with rate distortion theory (see for
example~\cite{csiszar:koerner}) makes it plausible
that by allowing a certain loss we can save even more in
the output entropy. This is because on blocks the fidelity
obeys the same form of rule as the typical distortion
measures: it is the average over the block.
\par
Several distortion criteria could be used, e.g.
$$F({\bf A})\geq F({\bf a})-d,$$
but many others seem natural, too.
\par
A similar tradeoff may occur between the optimum compression
rate and the parameter $g$, $\nu_l=\lfloor gl\rfloor$
in condition C5 (here we have treated only the case $g=0+$).
\par
We intend to pursue these questions in future work.

\acknowledgements
  This work was partially supported by the ESF,
  enabling discussions at the CCP workshop
  at Cambridge (July 1999) and two scientific visits to Brussels.
  SM is a research associate of the Belgian National Research Fund (FNRS).
  AW was supported by the SFB 343 ``Diskrete Strukturen in der Mathematik''
  of the Deutsche Forschungsgemeinschaft. SM acknowledges funding by
  the European Union  project EQUIP (contract  IST--1999--11053).


\end{document}